# Relaxation Oscillations and Ultrafast Emission Pulses in a Disordered Expanding Polariton Condensate

**Maciej Pieczarka**[1,*], **Marcin Syperek**[1], **Łukasz Dusanowski**[1,2], **Andrzej Opala**[1], **Fabian Langer**[2], **Christian Schneider**[2], **Sven Höfling**[2,3], **and Grzegorz Sęk**[1]

[1]Laboratory for Optical Spectroscopy of Nanostructures, Department of Experimental Physics, Faculty of Fundamental Problems of Technology, Wrocław University of Science and Technology, Wyb. Wyspiańskiego 27, 50-370 Wrocław, Poland

[2]Technische Physik, Physikalisches Institut and Wilhelm Conrad Röntgen-Research Center for Complex Material Systems, Universität Würzburg, Am Hubland, D-97074 Würzburg, Germany

[3]SUPA, School of Physics and Astronomy, University of St. Andrews, St. Andrews, KY 16 9SS, United Kingdom

## Abstract

Semiconductor microcavities are often influenced by structural imperfections, which can disturb the flow and dynamics of exciton-polariton condensates. Additionally, in exciton-polariton condensates there is a variety of dynamical scenarios and instabilities, owing to the properties of the incoherent excitonic reservoir. We investigate the dynamics of an exciton-polariton condensate which emerges in semiconductor microcavity subject to disorder, which determines its spatial and temporal behaviour. Our experimental data revealed complex burst-like time evolution under non-resonant optical pulsed excitation. The temporal patterns of the condensate emission result from the intrinsic disorder and are driven by properties of the excitonic reservoir, which decay in time much slower with respect to the polariton condensate lifetime. This feature entails a relaxation oscillation in polariton condensate formation, resulting in ultrafast emission pulses of coherent polariton field. The experimental data can be well reproduced by numerical simulations, where the condensate is coupled to the excitonic reservoir described by a set of rate equations. Theory suggests the existence of slow reservoir temporarily emptied by stimulated scattering to the condensate, generating ultrashort pulses of the condensate emission.

[*] e-mail: maciej.pieczarka@pwr.edu.pl

## Introduction

Condensates of exciton-polaritons[1,2], quasiparticles created from strong coupling between photons and excitons, have opened the way for studies of novel out of equilibrium physics. The polariton quasiparticle lifetime, despite long-living exciton in a quantum well, is typically short being in the range of a few to tens of picoseconds as inherited from the photon storage time within the microcavity[3–6]. Under these circumstances, the very complex dynamics of the condensate can be driven by the competition between gain and loss in the condensate, similarly to what is observed in photon lasers and other nonlinear systems. This open-dissipative nature of exciton-polaritons places them as a platform for realisation of non-Hermitian physics[7,8]. Moreover, the long-living excitonic reservoir not only feeds the condensate, but also acts as a repulsive potential for polaritons and therefore, determines the flow of polaritons in space[9] and time, and it serves as well as an additional decoherence source of the macroscopic state[10,11].

Polariton condensates were realised in a plethora of inorganic and organic material configurations[12–16]. Typically the semiconductor microcavities are not free from structural imperfections that introduce non-negligible disorder acting as a static external potential experienced by the nonlinear polaritonic waves[17]. The feature of scattering on defect potential led to observation of superfluid properties of polariton condensates[18,19]. Moreover, polariton localisation within structural defects plays a crucial role in coherent dynamics[11,20] of the system fed by hot excitons, showing features known from the physics of bosonic condensation [21,22] as well as effects originating from a dynamical imbalance of the reservoir and the condensate, where the combination of these generates new self-oscillatory effects[23]. Slow reservoir relaxation can also cause a dynamical instability in the formation process of an exciton polariton condensate[24,25]. In particular, nonequilibrium driven-dissipative condensates are much more influenced by disorder than their thermodynamic counterparts, where lack of the condensate stabilisation and superfluidity is predicted under any particle density[26–28].

In this work, we study nontrivial dynamics of a polariton condensate created in a semiconductor microcavity with significant disorder[29]. Repulsive interactions of polaritons and excitons create a localised potential within the focused laser spot that causes the expansion of the polariton wave, which subsequently interacts with a structural disorder. Contrary to the highly homogeneous semiconductor microcavity samples[30,31], here the polariton cloud is irregular reflecting the scattering features on defects. We analyse near-field photoluminescence patterns to investigate the temporal dynamics of the system in a time-resolved experiment. It reveals complex oscillatory behaviour at long propagation distances and densities above the condensation threshold. The emission bursts depend on the pump power densities. We interpret the creation of ultrashort pulses of polariton condensate emission in terms of slow reservoir relaxation oscillations. The excitonic reservoir is temporarily burnt out by the condensate and needs to be replenished by relaxing high-energy electron-hole pairs to reach once again the condensation threshold density. These results can be reconstructed well by theoretical simulations within the phenomenological mean field approach taking into account a set of rate equations describing the exciton formation dynamics in the quantum well. Theoretical simulations suggest that the pulsating dynamics are the natural state of the condensate emission for a parameter set tuned to the properties of investigated sample.

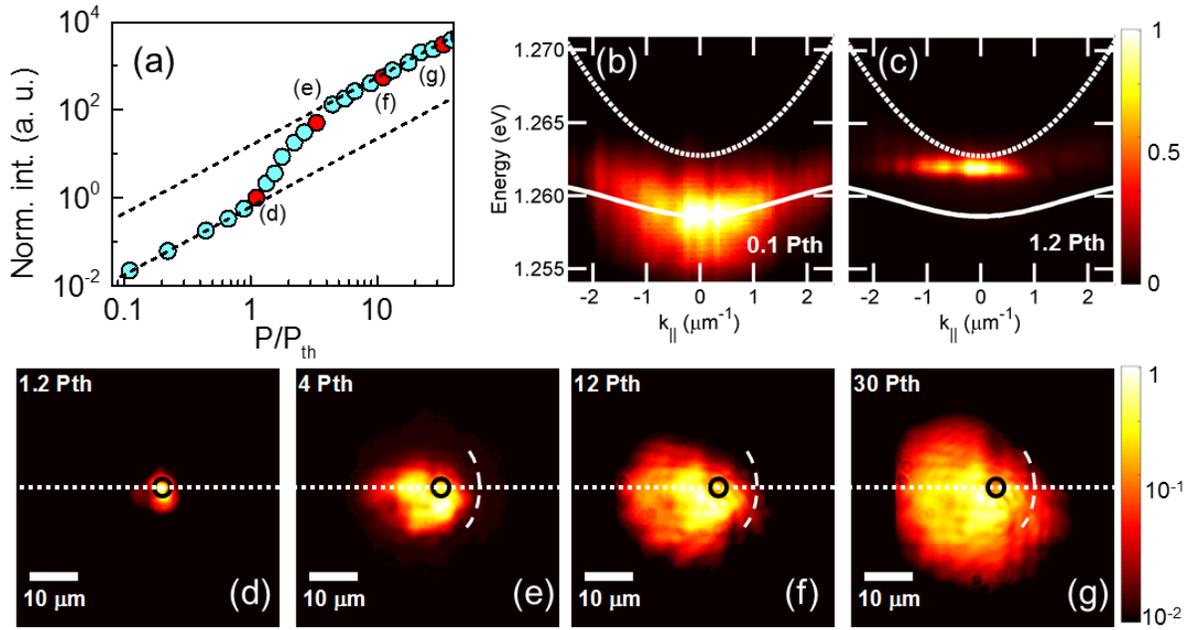

**Figure 1 Distribution of polariton condensate in a time-integrated picture.** (a) Power-dependent input-output characteristics of the polariton condensate emission. Power dependent k-space emission below (b) and above (c) the regime of polariton condensation. Bare cavity mode is depicted as dotted line and lower polariton branch as a white thick line. The bottom panel presents the real space distributions of polariton condensate at different excitation levels above the condensation threshold $P_{th}$, (d) 1.2 $P_{th}$, (e) 4 $P_{th}$, (f) 12 $P_{th}$ and (g) 30 $P_{th}$. The pseudo-color scale is logarithmic. The pump spot size and location is indicated as a black circle. The strong defect wall is indicated with a curved dashed line (guide to the eye). The system detection slit, the direction where the spectra are cut, is indicated as a dotted line.

## Results

### Time-resolved photoluminescence

In our experiment, we excited the system by a beam of laser pulses focused to a diffraction limited spot, which acts as a spatially localised source of the reservoir feeding the polariton condensate (see Methods). The absorbed photons create the initial population of electron-hole pairs, which relax via particle-particle and particle-phonon scattering, lowering their energy and forming high k-vector excitons with large energy. Excitons with small k-vectors inside the light cone are the main source of bosonic stimulation to the final state, achieving a hybrid light-matter polariton condensation (or polariton lasing) and driving the dynamics of the system. The power dependent input-output curve is shown in Fig. 1(a), presenting a characteristic nonlinear increase above the threshold density $P_{th}$. At low pumping conditions below the threshold, polariton emission follows the dispersion of lower polariton branch, Fig. 1(b). Above $P_{th}$ a spectrally narrowed mode is observed with significant blueshift due to interactions between polaritons and excitons, which is a typical signature of polariton condensation[32]. We further investigate the system in the above threshold conditions, where the density of the reservoir is high enough to provide gain for stimulated scattering into the final occupied mode which surpasses the loss of photons from the cavity. Additionally, the sample was previously characterized by spatially and spectrally resolved photoluminescence (PL) mapping to estimate the characteristic disorder distribution[29].

The spatial distribution of the polariton condensate is analysed in the time-integrated picture to get the knowledge on the behaviour of polariton propagation. Slightly above the condensation threshold ($P = 1.2P_{th}$), a submicron localized polariton cloud is observed in the close vicinity of the excitation spot as shown in Fig. 1(d). One can already observe a spatially irregular shape of the luminescence, which does not reflect the cylindrical symmetry of the focused laser (indicated as a black circle in Figs. 1). At weak optical excitation the density of the excitonic reservoir and the condensate is not high enough to generate energy blueshift from interactions between polaritons and the reservoir to generate the condensate outside the localizing potential in the vicinity of the excitation source. This is manifested in a flat dispersion at threshold, which is typically observed for a localized state[22], see Fig. 1(c). A different scenario takes place at higher excitation densities, when the polaritons overcome the local irregularities and expand to larger distances outside the pump spot, as seen in Fig. 1(e)-(g). At high excitation densities the polariton-polariton and polariton-exciton interactions within the sample cause a blueshift of around 2-3 meV, larger than the mean localization energy among structural disorder in the sample[29]. Thus, polaritons are expanding radially driven by the repulsion from the excitonic reservoir. However, one can observe that polariton propagation is blocked on the right hand side from the laser pump spot (Fig. 1 (e)-(g)), even at excitation conditions well above the threshold. It suggests the existence of a strong potential barrier that blocks the polariton flow in this direction. On the contrary, freely propagating polaritons at distances as large as 20 μm are observed on the left hand side of the map (being only slightly distorted by sample imperfections). Such extensive spread of the condensate cloud is a consequence of a ballistic flow of the coherent condensate wave.

At the arbitrarily chosen place on the sample, one observes a strong hampering of the polariton flow by a potential barrier in one direction and almost undisturbed propagation in the other direction. One should note that the time-integrated images are stable within the entire time of the experiment, which is a signature of similar polariton dynamics after each process of polaritons generation by the consecutive laser pulses. To reveal the true spatial dynamics of the propagation of polaritons and their interactions with disorder potential in the vicinity of the pump spot, we performed the time-resolved measurement using a streak camera. In this experiment, we could isolate the PL emission from the left-hand side and right-hand side positions of the pump spot area and analyse it in time-domain. Within this approach, one can record the dynamical behaviour of the condensate propagation which is hidden in time-integrated measurement.

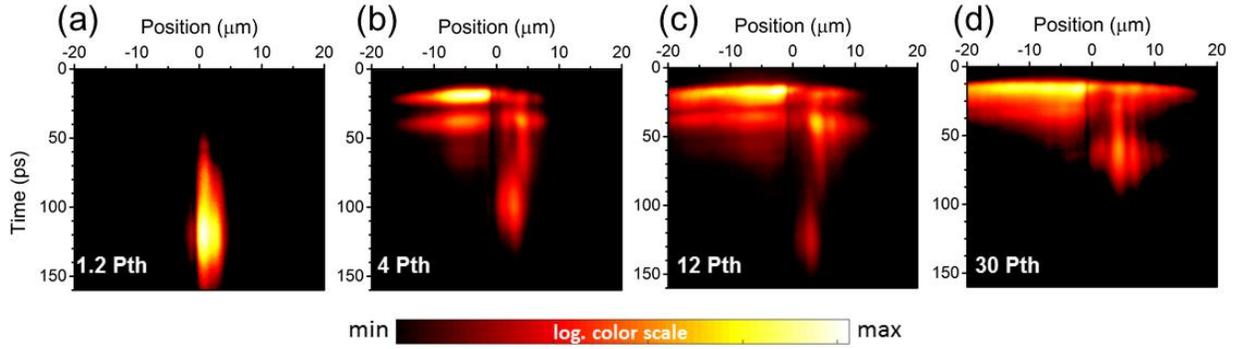

Figure 2 **Time-resolved spatial dynamics of a polariton condensate** Recorded-time resolved along the line indicated in Fig.1 are presented for several pumping powers. (a) 1.2 $P_{th}$, where long-lasting localized polariton emission is observed (b) 4 $P_{th}$, onset of polariton emission bursts, (c) 12 $P_{th}$ and (d) 30 $P_{th}$, merging of the pulses at highest pumping levels. The pseudo-color scale is logarithmic.

The obtained time-resolved PL maps are presented in Fig. 2 for spatial positions at the centre of the pump spot, according to the time-integrated images from Fig. 1, along with the indicated detection system slit. At $P \approx 1.2 P_{th}$ (Fig, 2 (a)), the related PL signal has a significant time delay and slow dynamics. This reflects mainly the electron-hole relaxation characteristics at lower densities around the condensation threshold, as the scattering from reservoir to the condensate mode is not dominated by bosonic stimulation[33]. As it was already mentioned, the emission originates from a condensate trapped within local potential minima, most probably related to a structural defect. At higher pumping powers, the recorded time-resolved spectra reveal very rich phenomenology. The emission rise time became extremely short, which is due to the enhanced bosonic stimulation at higher excitation densities above the threshold. Moreover, in Fig. 2(b) one can see freely propagating polariton waves on one side (negative spatial direction) and blocked polaritons with long-lasting irregular oscillations on the other side (positive spatial direction). The larger density of the excitonic reservoir the easier it is to overcome the local trapping potential by the propagating condensate. Consequently, the polariton condensate can be ballistically emitted outside the laser focal point, if not disturbed by the existence of some high energy potential barrier. Interestingly, the PL evolution consists of a bursts of polaritonic emission processes separated from each other by a few ps. It is a qualitatively different behavior in comparison to what has been observed in nearly defect-free, high structural quality polaritonic structures in the regime of ballistic dynamics[34,35], where polaritons decay in a single process. The observed pulsating dynamics sensibly depend on the pump power. As the excitation density increases, the burst pattern is gradually smearing over time. It transforms the polaritonic decay into a regime of a fast propagating wave that is clearly observed for the negative values at the position axis in the sequence of figures: Fig.2 (b), (c) and (d). The burst-like polaritonic emission pattern resembles the spatial-hole burning effect as observed in time evolution for polariton condensates in localized potential traps governed by slow reservoir relaxation[23]. In our case, one needs to take into account the ballistic propagation of emitted polaritons outside the pump spot.

The polaritons escape from the defect potential trap due to strong repulsive interactions and are emitted radially.

It is worth noting, that even for the largest pumping powers, the polaritons propagating in the direction of a high potential barrier are blocked and cannot propagate on macroscopic distances, as the waves in other direction, Fig. 2(d). This is an indication of an extremely large obstacle at this location, which amplitude is much larger than the observed blueshift. One can observe a modified dynamics due to the presence of such strong defects, changing the temporal behaviour different from the free propagating case.

**Theoretical simulations**

In order to get a better understanding of the observed dynamical properties, we performed theoretical simulations of polariton expansion under influence of defects, solving the Gross-Pitaevskii equation coupled to the incoherent reservoir rate equations[36]. It has been shown, that a proper description of time-resolved dynamics of an incoherently excited exciton-polariton condensate is only possible when one includes a multiple-step relaxation of the high-energy carriers before the reservoir feeds the condensate in a stimulated scattering process[37–39]. We have chosen the simplified model of carrier dynamics, including physical phenomena, which play an important role in the investigated structure. In a non-resonant excitation, photons from the laser pulse generate a population of free electrons and holes with high energy. We describe its spatial population $n_{eh}$ with a rate equation:

$$\frac{\partial n_{eh}(\mathbf{r},t)}{\partial t} = -\Gamma n_{eh}^2(\mathbf{r},t) - \gamma_{eh} n_{eh}(\mathbf{r},t) + P(\mathbf{r},t),$$

where $P$ is the spatial function of the pump pulse creating the initial density of $n_{eh}$. The uncorrelated electron-hole plasma can decay nonradiatively, which is described by the term $\gamma_{eh}$, or can form excitons with a rate $\Gamma$[3,40]. We assume, that the excitonic reservoir $n_R$ follows the second rate equation

$$\frac{\partial n_R(\mathbf{r},t)}{\partial t} = -R|\psi(\mathbf{r},t)|^2 - \gamma_R n_R(\mathbf{r},t) + \Gamma n_{eh}^2(\mathbf{r},t),$$

where $n_R$ can either feed the condensate field $\psi$ via stimulated scattering rate $R$ or decay in incoherent processes with the rate $\gamma_R$. The scalar polariton field is govern by a well-known mean-field Gross-Pitaevskii equation, taking into account interactions, gain and loss in the system

$$i\hbar \frac{\partial \psi(\mathbf{r},t)}{\partial t} = \left[ -\frac{\hbar^2 \nabla^2}{2m^*} + g_c|\psi(\mathbf{r},t)|^2 + g_R\big(n_R(\mathbf{r},t) + n_{eh}(\mathbf{r},t)\big) + V(\mathbf{r}) + \frac{i\hbar}{2}(Rn_R(\mathbf{r},t) - \gamma_c) \right] \psi(\mathbf{r},t).$$

Here, the polariton effective mass is given by $m^*$, decay rate $\gamma_c$, the mean-field polariton-polariton interaction constant $g_c$, and the interaction with reservoir is described with $g_R$. In the incoherent part of the system equations we neglect the diffusion of free carriers and excitons due to the fact, that electron and hole effective masses are several orders of magnitude larger than the effective mass of exciton-polaritons in the lower polariton branch $m^*$. The procedure of parameters choice is described in the Methods section. For qualitative simulation of the experiment, we chose a particular defect potential $V(r)$, based on experimental observations described above. The potential is composed of two Gaussian peaks of different amplitudes, where the smaller one is located at the left-hand side and an extreme defect at the right hand side, as indicated in the inset of Fig. 3, keeping its parameters reasonably close to real sample's potential[29]. In this approach we obtained a trapping potential in the middle, which captures the condensate within it, when the pumping level is around the condensation threshold. The obtained simulation results are presented in Fig. 3.

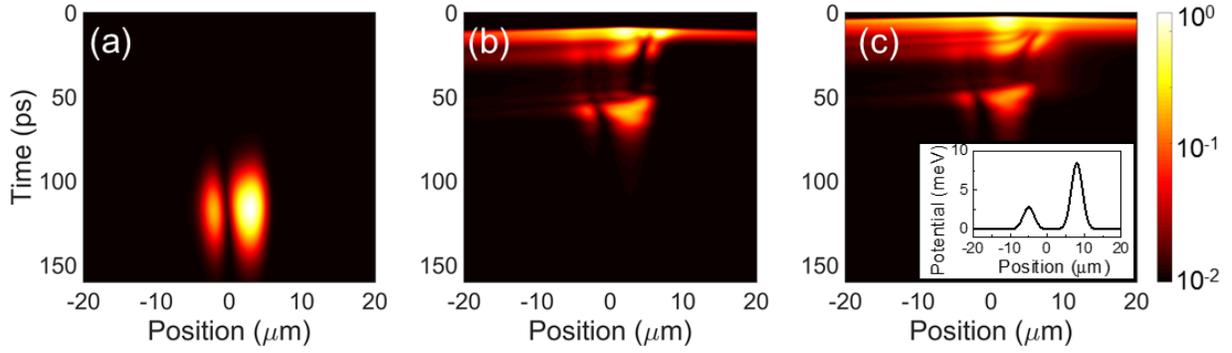

**Figure 3 Simulation results with disorder potential** Simulation results of a time-resolved propagation of polariton condensate after a pump pulse strength of (a) $1.5P_0$, (b) $8P_0$, (c) $15P_0$. $P_0$ stands for pumping amplitude to observe first condensation pulse. Simulation results are convolved with a streak camera system response function to emulate the experimental recordings. The pseudo-color scale is logarithmic and the intensities are normalized to the maximal value in each case. The chosen disorder potential is presented as an inset of (c).

One can observe that all characteristics of the experimental dynamics are qualitatively captured in the simulation. A long pulse located within the pump spot and appearing after a significant delay, comparable to recorded experimental dynamics in Fig. 2(a) is obtained near the condensation threshold, Fig. 3(a). At higher pumping rates, polaritons are ejected in the negative direction, where the local potential barrier is small. More importantly, the multiple pulse polariton dynamics are also observed, caused by reflection and trapping by the potential. On the other hand, a strong potential barrier in the positive part of the space blocks the flow and triggers the irregular oscillations, as seen both in the experiment and simulation. Finally, at higher pumping levels the simulation reproduces merging of multiple pulses into one fast polariton propagation of the polaritonic nonlinear wave in the negative direction.

The shape and time dependency of the complex dynamics are an extension of oscillatory behaviour in localised potentials[23], which was previously described in a semiclassical rate-equation model. Therefore, to elucidate if the particular shape of the localising potential is responsible for the observed time oscillations, we decided to simulate the dynamics using the same phenomenological parameters, but in absence of disorder potential. The results are presented in Fig. 4. Interestingly, the oscillatory dynamics are also obtained with all characteristic features like shortening of the oscillation period and merging of the propagating pulses. This suggests that the disorder traps are not the main cause of the particular condensation dynamics. It appears that for this set of system parameters the self-oscillation is the natural solution of the condensate dynamics within the model used. The reservoir dynamics are responsible for pulsating behaviour of condensate emission. The stimulated scattering above the threshold burns out the excitonic reservoir, which is replenished by relaxing high-energy electron-hole plasma. When the reservoir density is once again recovered above the critical level, another condensate pulses are emitted. The emission pulses disappear in time as the general density of photoexcited particles decays.

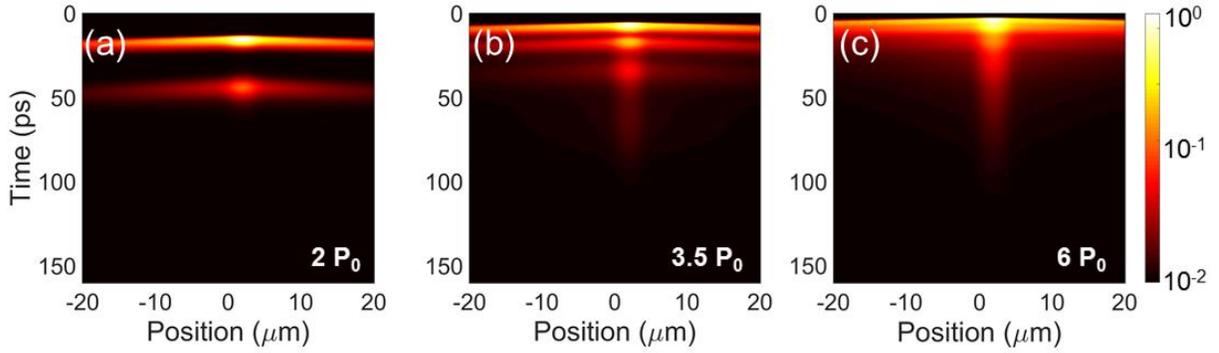

Figure 4 **Simulated dynamics of a free-propagating polariton condensate.** Simulation results for pumping strengths (a) $2P_0$, (b) $3.5P_0$, (c) $6P_0$. $P_0$ stands for pumping amplitude to observe first condensation pulse. The condensate parameters are the same as for the disordered case. Simulation results are also convolved with a streak camera response function. The pseudo-color scale is logarithmic and the intensities are normalized to the maximal intensity.

## Discussion

The oscillatory time evolution of exciton-polariton condensates has been observed and discussed in many contexts. Coupled condensates can form a Josephson-like junction, where they are caused by the dynamical Josephson effect[21,41]. Spin-dependent interactions of exciton polaritons can also cause polarisation beats[42]. However, this kind of oscillations origin from the coherent coupling of polariton condensates and interactions between them. On the other hand, the oscillatory behaviour of polariton condensate can also be driven by specific properties of the incoherent reservoir feeding the final state through many nontrivial physical mechanisms[43,44]. This scenario also holds in our case where one has to take into account the properties of incoherent reservoir relaxation[23] and long excitonic reservoir lifetime $\gamma_R = 1/\tau_R$. In the range of parameters, where the reservoir is much slower than the polariton condensate, it has been shown that the nonlinearity of the system can become effectively attractive[44,45]. This causes instabilities of the polariton condensate revealed in many dynamical forms[25,46], like relaxation oscillations and random condensate domain formation. We would like to emphasize that evaluation of a set of parameters describing the system was prepared so to minimize the number of fitting parameters in our simulations while keeping resonable description of the investigated system. However, it is impossible to erase the built-in disorder, which shapes the relaxation behaviour of the condensate, Figs. 3, 4. Thus, it is challenging to rule out the influence of disorder on the observed peculiar dynamics experimentally. However, it seems that the excitons in the high In content InGaAs quantum wells meet the conditions of a long living reservoir, for instance, due to strong exciton localization within the quantum wells fluctuations[47–49], which elongates the lifetime parameter. Quantum well exciton properties together with short lifetime of photon part, due to the moderate quality of the investigated microcavity, might result in the fast depletion of the reservoir and cause burst-like pulsating behavior of the emission process. Such specific coupling to the slowly relaxing reservoir was expected theoretically to enhance the signal of the ghost branch in Bogoliubov excitation spectrum[50] in a photoluminescence experiment, previously observed experimentally in the investigated sample[29]. This is an indirect indication of the long-living reservoir properties, which are described here.

Furthermore, the analytical solution of the described theoretical model for the excitonic reservoir density at the condensation threshold is $n_R^{th} = \gamma_c/R$. For low $\gamma_c$ the threshold density of the reservoir is also lower. In an above threshold conditions stimulated scattering can drop the reservoir density below critical level more easily (taking into account typical values of the scatterin rate $R$). However, when one assumes high quality samples with larger $\gamma_c$ (much larger than for the sample investigated here), the reservoir density becomes also larger and it is less possible to drain it below critical condensation level. This considerations suggest, that lower quality microcavities could be more favourable to observe oscillation relaxations in a polariton condensate formation and why such oscillatory behaviour is not usually observed in the state-of-the-art samples. Additionally, ballistic expansion of polaritonic waves outside of the laser source might have additional impact on faster emptying of the localised excitonic reservoir. All in all, the experimental configuration and the sample design needs to be established, where the self-pulsating dynamical instability is a natural state of the system. Finding such a solution in a cavity material design could bring polariton condensates as potential sources of ultra-short light pulses in the VCSEL configuration, without the neccesity of external cavity design[51,52].

To conclude, we investigated an extended exciton-polariton condensate, created within a moderate quality cavity with non-negligible disorder potential. Near-field images of polariton condensate luminescence exhibit an irregular spatial pattern, which was influenced by strong disorder traps. We tracked the time-resolved spatial dynamics of the system, which revealed the complex burst-like behaviour of polariton condensation. This was interpreted in a relaxation oscillation condition in the condensate formation. The experimental observations were successfully reproduced within the mean-field model, taking into account realistic parameters of the investigated system and the dynamics of exciton formation in quantum wells. Joint experimental and theoretical results suggest a slowly relaxing reservoir driving the pulsatory dynamics of the system.

## Methods

**Sample and experiment**

The semiconductor microcavity is composed of two distributed Bragg reflectors (DBRs). The upper DBR consists of 12 GaAs/AlGaAs mirror pairs, whereas the bottom one has 16 pairs. The cavity length is $\lambda/2$. At the antinodes of the confined optical mode, stacks of $In_{0.27}Ga_{0.73}As$ quantum wells (QWs) are located. The Rabi splitting of the polariton modes is about 7.5 meV. Experiments were carried out on the position on the sample, where the exciton-photon detuning was approximately $\Delta \approx 0$ meV. The sample is characterized with a built-in disorder, having the potential energy as large as 3 meV and broad spectral distribution, comparable to microcavities from II-VI group of semiconductors[29].

The sample was excited with a pulsed Ti: Sapphire laser with 76 MHz repetition frequency and pulse length of about 140 fs. The laser wavelength was tuned to one of the DBR reflectivity minima ~830 nm, close to the absorption band edge of GaAs, to optimize the injection efficiency of carriers in the structure, to prevent redundant absorption of the DBR material and to avoid unnecessary heating of the sample or change the refractive index of the cavity. The laser was focused on the sample surface to a spot of about 2.5 μm in diameter. Emission from polaritonic condensate was collected via high numerical aperture objective (NA=0.42) and imaged (real or Fourier space) by a set of lenses on the manually adjusted slit of the detection system. The slit was varied with a micrometre precision. The detection system is composed of a half-meter-long monochromator coupled to a streak camera equipped with the S1 photocathode. The overall temporal resolution of the system is around 3 ps (without the spectral dispersion). The scattered pump laser was filtered out by a set of high-quality edge filters of OD>6 around the laser wavelength.

**Simulation details and parameters**

The simulation was performed in a 1D approximation. Within this approach, the model does not capture nonradial propagation of polaritonic waves, but it is sufficient to describe oscillation relaxation in condensate formation and scattering on a large defect wall. The GPE was evaluated numerically with a split-step spectral method[53] and the reservoir rate equations were solved in a 4-th order Runge-Kutta method.

The exciton properties confined in the QW were numerically calculated employing the Nextnano software[54] (www.nextnano.com). Initially, single particle states in the conduction and valence bands were assessed within the effective mass approach. Then, a Coulomb interaction correction was included based on the single particle states as an input and the exciton binding energy and Bohr radius were obtained $E_b = 8.6$ meV, $a_B = 11.6$ nm, respectively. Based on theoretical estimation[55] $g_R \approx 6E_b a_B^2 = 6.8\ \mu\text{eV}/\mu\text{m}^2$ and for polariton-polariton interaction we chose $g_c = |X|^2 g_R$, where $|X|^2$ is the Hopfield coefficient. The reservoir decay time was measured below the polariton lasing threshold in the investigated system, where single-exponential decay yields $\gamma_R = 1/80\ \text{ps}^{-1}$. Scattering of free electron-hole pairs to the excitonic reservoir is modelled based on literature parameters[3,37,39,40], and the used parameters are $\Gamma = 10^{-4}\ \mu\text{m/ps}$ and $\gamma_{eh} = 1\ \text{ns}^{-1}$. The photon lifetime was calculated based on the Q-factor measurements of the sample and taking into account the Hopfield coefficient $\gamma_c = 0.33\ \text{ps}^{-1}$. The stimulated scattering rate was tuned to obtain the experimentally observed blueshift at the condensation threshold $\hbar R = 8.0\ \mu\text{eV/ps}$. To simulate the pulsed excitation, we used a $\delta$-like pump pulse in time, which sets the initial spatial population of free electron-hole pairs $n_{eh}(t = 0)$ according to the pumping profile.

**Data availability**

The data that support the plots within this paper and other findings of this study are available from the corresponding author upon reasonable request.

**Acknowledgments**

Authors would like to acknowledge inspiring discussions with Michal Matuszewski.

## Author contributions statement

M.P., M.S., Ł.D. performed the experiments, M.P. analyzed the experimental data. M.P. and A.O. performed the theoretical simulations. F.L., C.S. and S.H. fabricated the sample. G.S. coordinated the research project. All authors discussed the results and contributed to the manuscript preparation.

## Additional Information

### Competing financial interests

The authors declare that they have no competing interests.